\documentclass[twocolumn,showpacs,preprintnumbers,amsmath,amssymb,aps,prb,floatfix,superscriptaddress]{revtex4}

\usepackage{graphicx}
\usepackage{dcolumn}
\usepackage{bm}

\input colordvi

\begin{document}

\title{Computationally-Driven Experimental Discovery of the CeIr$_4$In Compound}
\author{D. J. Fredeman}
\affiliation{Department of Materials Science and Engineering,
                  Cornell University, Ithaca, New York 14853, U.S.A.}
\affiliation{Theoretical Division and Center for Nonlinear Studies,
                  Los Alamos National Laboratory, Los Alamos, New Mexico 87545, USA.}
\author{P. H. Tobash}
\affiliation{Materials Physics \& Applications Division, Los Alamos National Laboratory,
                   Los Alamos, New Mexico 87545, U.S.A.}
\author{M. A. Torrez}
\affiliation{Materials Physics \& Applications Division, Los Alamos National Laboratory,
                   Los Alamos, New Mexico 87545, U.S.A.}
\author{J. D. Thompson}
\affiliation{Materials Physics \& Applications Division, Los Alamos National Laboratory,
                   Los Alamos, New Mexico 87545, U.S.A.}
\author{E. D. Bauer}
\affiliation{Materials Physics \& Applications Division, Los Alamos National Laboratory,
                   Los Alamos, New Mexico 87545, U.S.A.}
\author{F. Ronning}
\affiliation{Materials Physics \& Applications Division, Los Alamos National Laboratory,
                   Los Alamos, New Mexico 87545, U.S.A.}
\author{W. Tipton}
\affiliation{Department of Materials Science and Engineering,
                  Cornell University, Ithaca, New York 14853, U.S.A.}
\author{Sven P. Rudin}
\affiliation{Theoretical Division, Los Alamos National Laboratory,
                   Los Alamos, New Mexico 87545, U.S.A.}
\author{R. G. Hennig}
\email{rhennig@cornell.edu}
\affiliation{Department of Materials Science and Engineering,
                  Cornell University, Ithaca, New York 14853, U.S.A.}
\date{\today }

\begin{abstract}
We present a combined experimental and computational methodology for the discovery of new materials. Density functional theory (DFT) formation energy calculations allow us to predict the stability of various hypothetical structures. We demonstrate this approach by computationally predicting the Ce-Ir-In ternary phase diagram. We predict previously-unknown compounds CeIr$_4$In and Ce$_2$Ir$_2$In to be stable. Subsequently, we successfully synthesize CeIr$_4$In and characterize it by X-ray diffraction. Magnetization and heat capacity measurements of CeIr$_4$In are reported. The correct prediction and discovery of CeIr$_4$In validates this approach for discovering new materials.
\end{abstract}



\maketitle

\section{Introduction}

The holy grail of computational materials design is the \emph{ab initio} discovery of new materials which can be synthesized to have specific desired properties. Our increasing understanding of the structure/properties relationship in many materials and our increasing computational resources make this an exciting possibility.~\cite{Piquini08} However, in many cases, the precise relationship between structure and properties is not yet well understood. This is the case for strongly correlated electron materials which exhibit interesting phenomena such as superconductivity, colossal magnetoresistance, and multiferroic behavior.~\cite{SCESDesign}

Despite lacking a theory to immediately predict the properties of strongly correlated materials from structure alone, phenomenological observations often suggest families of composition and types of structures which are likely to exhibit specific functionalities. Having identified a family of such structures, the question becomes whether one can synthesize new materials in these families having enhanced properties. Unfortunately, the number of candidate materials which one might identify is often very large, and the experimental costs in searching for compounds can thus be heavy. Therefore, a computational method for predicting the stability of hypothetical compounds is highly desirable.

A variety of computational structure search methods have been presented including random search methods, genetic algorithms and other heuristic methods, and sophisticated data-mining approaches.~\cite{Oganov10, Levy10, Fischer06, Walle02} However, such methods are non-ideal for our application; because we are interested in searching for structures at many different compositions, random search and heuristic methods often require a prohibitively large amount of computation time.  Formal data-mining approaches often fail in ternary and higher-order systems as there is an insufficient amount of data available.  In addition, all of these methods require a significant amount of resources devoted to software development.  In our case, the parallel progress of experiment and theory led to fast results.  Neither theory nor experiment alone could have progressed to explore this material system with modest devoted resources and on the same timescale of several months.

In this paper we report on our exploration of the Ce-Ir-In ternary phase diagram. Cerium-based transition metal indide compounds adopt a large variety of structure and bonding arrangements which produce a large variety of interesting physical phenomena including superconductivity.~\cite{Kalychak05} Two types of structures which we are interested in exploring include systems with Ce sublattices which exist in either square planar or geometrically frustrated arrangements. Experience with cuprates, iron-pnictide, organic, and heavy fermion superconductors suggests that the square-planar compounds are likely to exhibit superconductivity.~\cite{MonthouxNature07} In geometrically frustrated systems, large degeneracies can lead to novel physical phenomena.~\cite{RamirezReview94}

Using this method, we examine the structural stability of selected compounds in the Ce-Ir-In ternary phase diagram and predict two previously unknown compounds, CeIr$_4$In and Ce$_2$Ir$_2$In, to be stable. Subsequently, we were able to synthesize CeIr$_4$In, which possesses a geometrically frustrated Ce sublattice.

\section{Methodology}

\subsection{Thermodynamics} 

At thermal equilibrium, a compound is stable if it has a negative formation energy with respect to its constituent elemental and any other competing phases.~\cite{chang10}  The formation energy per atom for a ternary compound Ce$_n$Ir$_m$In$_l$ with respect to its constituent elements may be written as
\begin{equation}
\label{eq:eform}
 \Delta E_\mathrm{f} = \frac{E_{\textrm{Ce}_n\textrm{Ir}_m\textrm{In}_l } - n\ E_\mathrm{Ce} - m E_\mathrm{Ir} - l E_\mathrm{In}}{n+m+l} .
\end{equation}
where $E_\mathrm{Ce}$, $E_\mathrm{Ir}$, and $E_\mathrm{In}$ are the Gibbs free energies per atom of the elemental phases.

\begin{figure}[tb]
  \begin{center}
    \includegraphics*[width=8.5cm]{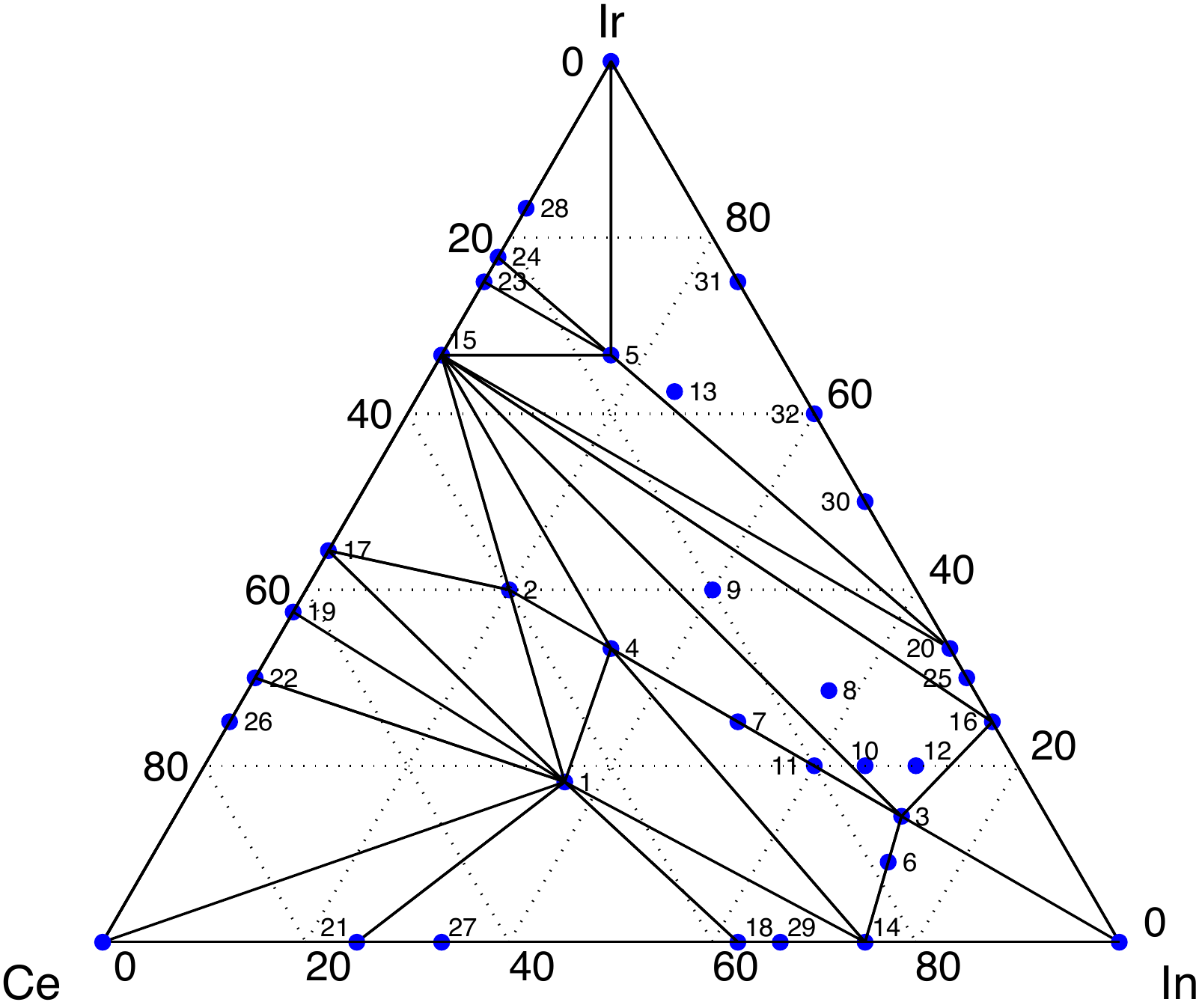}
    \caption{\label{fig:phasedia} (color online) Predicted Ce-Ir-In phase diagram.  The top-down view of the convex hull of the energy of the various Ce-Ir-In phases illustrates which phases are stable and unstable at zero temperature.  The ternary phases at corners of triangle represent stable ternary compounds, {\it i.e.} compounds such as Ce$_{5}$Ir$_{2}$In$_{4}$ (1), CeIrIn$_{5}$ (3), and CeIr$_{4}$In (5) are predicted to be stable. Compounds such as CeIrIn$_{2}$ (7), CeIr$_{2}$In$_{7}$ (12), and CeIr$_{5}$In$_{2}$ (13) fall within the hull, rather than on its surface, and are predicted to be unstable.  The binary and ternary compounds are numbered following their order in Table~\ref{tab:results}.}
  \end{center}
\end{figure}

Using these formation energies, stable and competing phases can easily be illustrated.  As shown in Fig.~\ref{fig:phasedia}, each compound is plotted according to its composition and formation energy.  Then, the smallest surface encompassing all points is known as the convex hull, and points which lie on the convex hull itself are thermodynamically stable.

Both our computational and experimental techniques have limitations in finding these low-energy structures.  Identifying candidate structures is an open challenge to the computational methods which also have known systematic errors.  The annealing process in experimental synthesis techniques may fail to find the lowest energy structure due to kinetic barriers and insufficient driving forces.  Thus, synthesis routes may exist which allow us to produce metastable compounds and at the same time it may be difficult to realize the true ground state.

\subsection{Details of the Calculations}

Calculation of the entire Gibbs free energy is computationally expensive and unnecessary.  The internal energy contribution to the free energy dominates the enthalpic and entropic terms, particularly at low temperature and pressure.  In this work, energies are calculated and evaluated in the zero temperature and pressure limit, neglecting zero point energies which are small in these heavy elements.

Our DFT calculations are performed using {\sc VASP}~\cite{kresse96, kresse99} with the projector-augmented wave PAW method in the generalized gradient approximation PBE.~\cite{blochl94a} Energy differences between compounds were converged to within 0.3~meV/atom using a plane wave cutoff energy of 400~eV and a k-point mesh of density 40~\AA$^{-1}$.  During the structural optimization, the Brillioun zone integration was performed using the Methfessel-Paxton scheme with a smearing of 0.2~eV and for the final structure, the energy was calculated using the linear tetrahedron method with Bl\"{o}chl corrections.~\cite{blochl94}

Our method of exploring the ternary phase space requires initial guesses of the crystal structure for a variety of compositions.  The results presented here take advantage of the fact that a compound's structure is often reasonably close to that of a related ternary compound composed of different elements.~\cite{Hennig05} For example, the structure of CeRhIn$_5$ can be used as a guess for the structure of CeTMIn$_5$, where TM is any transition metal.  Following this approach we assembled a set of candidate structures from known compounds in the Ce-Ir-In systems and related binary and ternary systems.  Specifically, we selected the known elemental, binary and ternary compounds of the Ce-Ir-In, Ce-Pt-In, Ce-Rh-In, and Ce-Pd-Al  ternary and the Ru-In binary systems.  We did not, however, include all known compounds of these systems, but rather selected those which experimental experience has shown most often form in Ce-Ir-In ternary phases and other chemically-similar systems.

After determining the formation energy for all the candidate structures, we determine the 3-dimensional convex hull, by plotting the composition versus formation energy with respect to the elemental phases.  Those compounds that form the convex hull are energetically stable while the compounds above the hull have a positive formation energy against competing phases and are unstable and hence difficult or impossible to synthesize.  From the hull, we can also identify competing phases. Calculating the formation energy of a compound with respect to its competing phases allows us to identify its degree of stability.  Stable compounds will exhibit a negative formation energy with respect to competing phases.

The success of this project is due to the close communication between experimental and theoretical efforts.  The sharing of preliminary information between the groups about particular structures and entire portions of the phase diagram was instrumental in accelerating the discovery process.

\subsection{Details of the Synthesis}

Compounds can be synthesized in a large variety of ways. For those compounds which were previously unknown to our knowledge (namely CeIr$_4$In and Ce$_2$Ir$_2$In) we attempted to synthesize these crystals by arc melting. The synthesis of CeIr$_4$In was carried out via arc-melting the respective elements in a stoichiometric ratio of Ce:Ir:In - 1:4:1.  The resulting button was flipped over 3 times and completely melted in order to ensure the reaction product was homogeneous.  After the arc-melting, the pellet was weighed and was found to have less than a 5~mg loss.  X-ray powder diffraction on the as cast pellet was completed on an X-1 Scintag diffractometer in the 2$\theta$ range of 10 - 80$^\circ$. The powder pattern was refined using the JADE software package. The attempted synthesis of Ce$_2$Ir$_2$In was carried out via arc-melting the respective elements in a stoichiometric ratio of Ce:Ir:In of 2:2:1. Characterization measurements of the magnetic susceptibility and heat capacity were performed in commercial systems from Quantum Design.

\begin{table*}
  \caption{\label{tab:results} Calculated energies, competing phases, and space groups associated with Ce-Ir-In phase diagram.  The total energy of the compound, $E_\mathrm{total}$, represents the calculated value per formula unit.  The heat of formation of the compounds, $\Delta E_\mathrm{f}$, relative to the elements and the formation energy relative to the competing phases, $\Delta E_\mathrm{comp}$,  are given units of meV/atom. Compounds listed in {\bf bold} are known to exist.}
  \begin{ruledtabular}
    \begin{tabular}{l l  d  c  c  c l}
      No. & Compound&
      \multicolumn{1}{c}{\hfil$E_\mathrm{total}$\hfil} &
      $\Delta E_\mathrm{f}$ &
      $\Delta E_\mathrm{comp}$ &
      Competing Phases &
      Space\ Group \\
      &
      \multicolumn{1}{c}{\hfil [eV/f.u.] \hfil} &
      \multicolumn{1}{c}{\hfil [meV/atom] \hfil} &
      \multicolumn{1}{c}{\hfil [meV/atom] \hfil} &
      &
      \\
      \hline
      1 & {\bf Ce$_{5}$Ir$_{2}$In$_{4}$} & -64.081 & -588 & -65 & Ce$_{3}$In$_{5}$, Ce$_{5}$Ir$_{3}$, Ce$_{2}$Ir$_{2}$In & $Pbam$ (55) \\
      2 & Ce$_{2}$Ir$_{2}$In & -35.881 & -749 & -42 & Ce$_{5}$Ir$_{4}$, CeIr$_{2}$, CeIrIn & $ P4/mbm$ (127) \\
      3 & {\bf CeIrIn$_{5}$} & -30.347 & -395 & -35 & Ce$_{2}$IrIn$_{8}$, CeIrIn, IrIn$_{3}$ & $P4/mmm$ (123) \\
      4 & {\bf CeIrIn} & -19.415 & -689 & -32 & Ce$_{2}$Ir$_{2}$In, CeIrIn$_{5}$  & $P\bar{6}2m$ (189) \\
      5 & {\bf CeIr$_{4}$In} & -46.651 & -455 & -8 & Ce$_{2}$Ir$_{7}$, IrIn$_{2}$ & $I\bar{4}$ (82) \\
      6 & {\bf Ce$_{2}$IrIn$_{8}$} & -45.594 & -400 & 1 & CeIn$_{3}$, CeIrIn$_{5}$ & $P4/mmm$ (123) \\
      7 & {\bf CeIrIn$_{2}$} & -21.987 & -520 & 40 & CeIrIn, CeIrIn$_{5}$ & $Cmcm$ (63) \\
      8 & CeIr$_{2}$In$_{4}$ & -36.504 & -375 & 44 & CeIr$_{2}$, IrIn$_{3}$, CeIrIn$_{5}$ & $Pmna$ (62) \\
      9 & CeIr$_{2}$In$_{2}$ & -31.124 & -472 & 56 & CeIr$_{2}$, IrIn$_{3}$, CeIrIn$_{5}$ & $P2\bar{1}/m$ (11) \\
      10 & Ce$_{3}$Ir$_{4}$In$_{13}$ & -93.408 & -346 & 70 & CeIr$_{2}$, IrIn$_{3}$, CeIrIn$_{5}$ & $Pm3m$ (223) \\
      11 & CeIrIn$_{3}$ & -24.435 & -394 & 89 & CeIrIn, CeIrIn$_{5}$ & $Pm3n$ (223) \\
      12 & CeIr$_{2}$In$_{7}$ & -43.520 & -197 & 140 & IrIn$_{3}$, CeIr$_{2}$, CeIrIn$_{5}$ & $I4/mmm$ (139) \\
      13 & CeIr$_{5}$In$_{2}$ & -56.784 & -181 & 189 & Ce$_{2}$Ir$_{7}$, Ir, IrIn$_{2}$ & $P4/mmm$ (123) \\
      \hline
      14 & {\bf CeIn$_{3}$} & -15.255 & -412 & -157 & In, Ce$_{3}$In$_{5}$ & $Pm3m$ (221) \\
      15 & {\bf CeIr$_{2}$} & -25.994 & -782 & -149 & Ce$_{5}$Ir$_{4}$, CeIr$_{3}$ & $Fd\bar{3}m$ (227) \\
      16 & {\bf IrIn$_{3}$} & -17.280 & -187 & -43 & In, Ir$_{3}$In$_{7}$ & $P4_2/mnm$ (134) \\
      17 & {\bf Ce$_{5}$Ir$_{4}$} & -71.542 & -717 & -43 & Ce$_{5}$Ir$_{3}$, CeIr$_{2}$ & $Pnma$ (62) \\
      18 & {\bf Ce$_{3}$In$_{5}$} & -33.643 & -382 & -30 & Ce$_{3}$In, CeIn$_{3}$ & $Cmcm$ (63) \\
      19 & {\bf Ce$_{5}$Ir$_{3}$} & -61.358 & -640 & -18 & Ce$_{7}$Ir$_{3}$, Ce$_{5}$Ir$_{4}$ & $P4/ncc$ (130) \\
      20 & {\bf IrIn$_{2}$} & -14.526 & -184 & -18 & Ir, IrIn$_{3}$ & $Fddd$ (70) \\
      21 & {\bf Ce$_{3}$In} & -21.036 & -170 & -17 & Ce, Ce$_{3}$In$_{5}$ & $Pm3m$ (221) \\
      22 & {\bf Ce$_{7}$Ir$_{3}$} & -73.297 & -520 & -7 & Ce, Ce$_{5}$Ir$_{3}$ & $P6_3mc$ (186) \\
      23 & {\bf CeIr$_{3}$} & -34.913 & -602 & -5 & Ce$_{2}$Ir$_{7}$, CeIr$_{2}$ & $R\bar{3}m$ (166) \\      
      24 & {\bf Ce$_{2}$Ir$_{7}$} & -78.684 & -535 & 0 & Ir, CeIr$_{3}$ & $R\bar{3}m$ (166) \\
      25 & Ir$_{3}$In$_{7}$ & -46.201 & -172 & 13 & IrIn$_{2}$, IrIn$_{3}$ &  $Im3m$ (229) \\
      26 & {\bf Ce$_{3}$Ir} & -28.252 & -399 & 34 & Ce, Ce$_{7}$Ir$_{3}$ & $Pnma$ (62) \\
      27 & {\bf Ce$_{2}$In} & -14.929 & -168 & 49 & Ce$_{3}$In, Ce$_{3}$In$_{5}$ & $P6\bar{3}/mmc$ (194) \\
      28 & {\bf CeIr$_{5}$} & -52.303 & -347 & 54 & Ce$_{2}$Ir$_{7}$, Ir & $P6/mmm$ (191) \\
      29 & {\bf CeIn$_{2}$} & -12.050 & -334 & 58 & Ce$_{3}$In$_{5}$, CeIn$_{3}$ & $Imma$ (74) \\
      30 & IrIn & -11.532 & -58 & 80 & Ir, IrIn$_{2}$ & $C2/m$ (12) \\
      31 & Ir$_{3}$In & -28.895 & 59 & 128 & Ir, IrIn$_{2}$ & $Pm\bar{3}m$ (221) \\
      32 & Ir$_{3}$In$_{2}$ & -31.479 & 42 & 152 & Ir, IrIn$_{2}$ & $P\bar{3}1c$ (163) \\
      \hline
      & Ce & -5.933 & & & & $Fm\bar{3}m$ (225) \\
      & Ir & -8.857 & & & & $Fm\bar{3}m$ (225) \\
      & In & -2.558 & & & & $I4/mmm$ (139) \\
    \end{tabular}
 \end{ruledtabular}
\end{table*}

\section{Results}
\subsection{Theory}

\begin{figure}[tb]
  \begin{center}
    \includegraphics*[width=8cm]{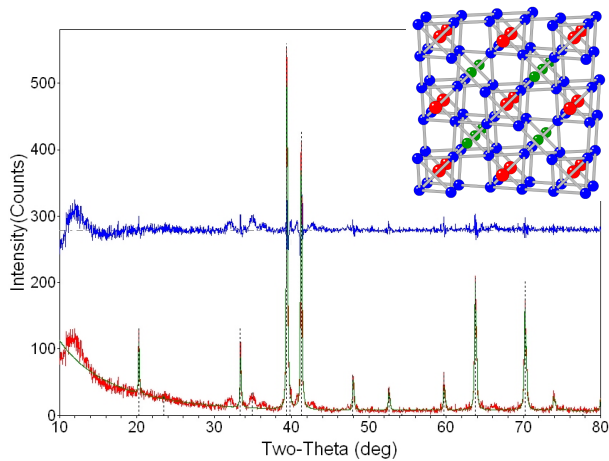}
    \caption{\label{fig:Xray} (color online) X-ray powder diffraction of CeIr$_4$In. Tick marks indicate the expected peak positions based on the MgCu$_4$Sn structure type. The difference between the measured (red) and calculated (green) spectrum is shown above the measured spectrum in blue. The inset displays the crystal structure of CeIr$_4$In (red spheres = Ce, blue spheres = Ir, and green spheres = In).}
  \end{center}
\end{figure}

For the Ce-Ir-In phase diagram we computed the formation energy of five known ternary structures (Ce$_{5}$Ir$_{2}$In$_{4}$\cite{Zaremba08Ce5Ir2In4}, CeIrIn$_{5}$\cite{CeIrIn5, ThompsonJMMM01}, CeIrIn\cite{ZarembaZAAC05}, Ce$_{2}$IrIn$_{8}$\cite{ThompsonJMMM01}, and CeIrIn$_{2}$\cite{ZarembaZAAC05}), eight currently unknown ternary structures (Ce$_{2}$Ir$_{2}$In, CeIr$_{4}$In, CeIr$_{2}$In$_{4}$, CeIr$_{2}$In$_{2}$, CeIrIn$_{3}$, Ce$_{3}$Ir$_{4}$In$_{13}$, CeIr$_{2}$In$_{7}$, and CeIr$_{5}$In$_{2}$), nineteen binary structures, and the three elemental phases. The results are tabulated in Table~\ref{tab:results}, and the corresponding convex hull is presented in Fig.~\ref{fig:phasedia}. Of the five known ternary compounds, three of them were computed to be stable. The two others, Ce$_{2}$IrIn$_{8}$ and CeIrIn$_{2}$, had only small positive formation energies of 0.7 and 40.2~meV with respect to competing phases, respectively. 
Of the eight unknown ternary structures, six have positive formation energies relative to competing phases, and thus we would expect them to be difficult to synthesize. However, two of these compounds have negative formation energies with respect to competing phases. Consequently we predict two new stable compounds: Ce$_{2}$Ir$_{2}$In and CeIr$_{4}$In.  Subsequent experiments described in Sec.~\ref{sec:experiments} successfully synthesized the CeIr$_{4}$In.

We characterize the electronic, magnetic and vibrational properties of the new CeIr$_{4}$In compound.   Phonon calculations for the cubic CeIr$_{4}$In compound show an unstable mode and predict a rotational distortion of the Ir tetrahedra.  Relaxing the crystal structure along this distortion mode rotates the tetrahedra by approximately 11$^\circ$ and couples to a tetragonal distortion of the lattice parameters elongating the lattice parameter parallel to the axis of rotation.  The space group symmetry is lowered from cubic $F\bar 43m$ (216) to tetragonal $I\bar 4$ (82).  The lattice parameter change slightly from $a = 7.566$~\AA\ of the cubic structures to $a=b=7.508$~\AA\ and $c=7.640$~\AA.  The symmetry breaking lowers the energy by about 300 meV per unit cell.   The X-ray diffraction pattern in Fig.~\ref{fig:Xray} shows no evidence of the cubic cell undergoing any such distortion. This discrepancy between calculations and experiment may potentially be due to long-range effects not captured in the limited cell size of the calculations, {\it e.g.}, disorder among the Ce and Ir sublattices. 

Calculations of the electronic and magnetic structure of the distorted CeIr$_{4}$In compound starting from various potential magnetically ordered states always result in a non-magnetic ground states.  Each Ce atom contributes one {\it f}-electron to the bonding.  The {\it f}-electrons is delocalized and no magnetic moments are present.  This suggests that the magnetic features seen in the experiments described in Sec.~\ref{sec:experiments} are indeed a consequence of a small amount of impurity phases.

\subsection{Experiment}
\label{sec:experiments}

Following the theoretical prediction of stable Ce$_{2}$Ir$_{2}$In and CeIr$_{4}$In compounds, we attempted to synthesize these compounds by arc melting.  The analysis of the X-ray powder diffraction data of Ce$_{2}$Ir$_{2}$In indicates multiple competing phases which are not easily identifiable.  Annealing of the Ce$_{2}$Ir$_{2}$In sample at 800$^\circ$C for two weeks did not significantly alter the powder X-ray diffraction pattern. In contrast, the X-ray analysis of the arc melted CeIr$_{4}$In sample shown in Fig.~\ref{fig:Xray} confirms the formation of the predicted CeIr$_{4}$In compound with lattice parameters $a = b = c = 7.579$~\AA\ in good agrement with the calculated lattice parameters.

The crystal structure of CeIr$_{4}$In is presented in the inset of Fig.~\ref{fig:Xray}. The Ce sublattice exhibits an fcc structure composed of corner sharing tetrahedra, which satisfies the condition for geometric frustration of the magnetic moments on the Ce atoms. Consequently, we measured the magnetic susceptibility and heat capacity to determine the basic properties and whether there is evidence for magnetic frustration in CeIr$_{4}$In.

\begin{figure}[tb]
  \begin{center}
    \includegraphics*[width=8cm]{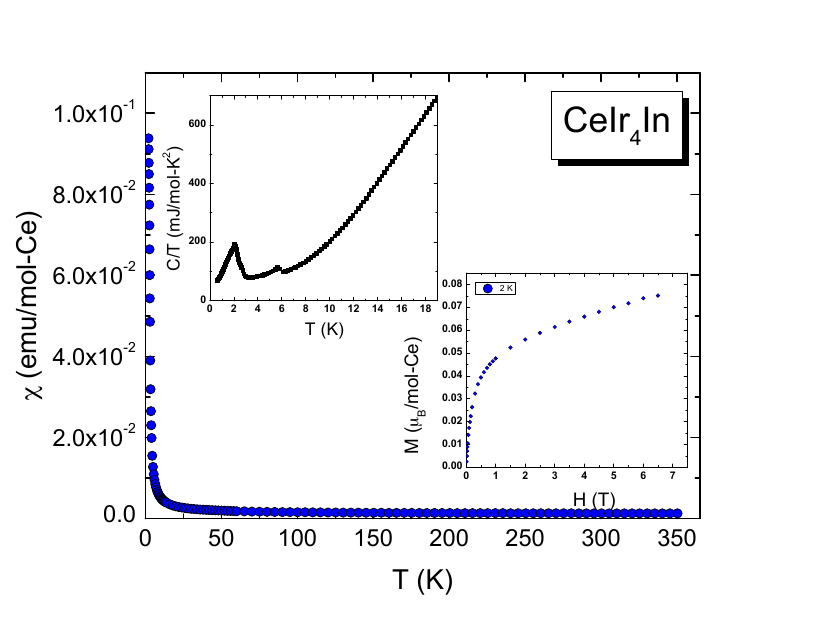}
    \caption{\label{fig:susc} (color online) Magnetic susceptibility of CeIr$_4$In versus temperature.  Upper left inset displays the specific heat of CeIr$_4$In versus temperature and the lower right inset shows the magnetization as a function of magnetic field at 2 K.}
 \end{center}
\end{figure}

The magnetic susceptibility is shown in Fig.~\ref{fig:susc}. An indication for ferromagnetic ordering at low temperature is indicated by the rapid rise in the susceptibility. The inset showing $M$ vs.\ $H$ further supports ferromagnetic ordering at low temperature albeit with a small ordered moment. The inverse susceptibility shows a large region of curvature up to room temperature. Similar temperature dependence seen in SmPt$_4$In was interpreted in terms of a Van Vleck contribution plus crystal electric field (CEF) effects.~\cite{MalikJMMM90} However, the CEF ground state for Ce in CeIr$_4$In is most likely magnetic. Alternatively, the large region of curvature could indicate a relatively large energy scale for the onset of exchange interactions, for which the ordering between moments is suppressed to much lower temperatures.

The heat capacity data shown in the inset of Fig.~\ref{fig:susc} shows a clear albeit broad magnetic ordering transition at 2.5~K consistent with the susceptibility data and resistivity measurements (not shown) and a smaller anomaly at 6~K. The field dependence of the anomaly at 2.5~K (not shown) and the shape of the transition in susceptibility again suggests that the magnetic ordering is ferromagnetic.  The extrapolated zero temperature Sommerfeld coefficient in the ordered state is 55~mJ/(mol$\cdot$K$^2$).  From the specific heat data, we calculate the entropy $S=\int C/T\, dT$.  However, only about 5\% of $R\ln 2$ in entropy is recovered at the transition temperature, much smaller than expected for a system of local magnetic moments at the ferromagnetic ordering temperature.

The small amount of entropy and small magnetic moment inferred from the susceptibility measurements suggests that these features are a consequence of a small amount of impurity phases. We note that there are small unaccounted peaks in the X-ray pattern shown in Fig.~2. While we believe this is the most likely explanation a few alternative scenarios still exist. First, the reduced entropy may be a consequence of Ce moments having largely itinerant character ({\it i.e.} in a Ce$^{4+}$ configuration) as has been suggested to occur in CeNi$_4$In.~\cite{TsujiiPhysicaB03} Alternatively, a large majority of the magnetic entropy may be quenched at much higher temperatures as a result of disorder-, RKKY-, and/or geometric-induced magnetic frustration effects as has been suggested to occur in the $R$Cu$_4$In ($R$ = Gd-Er) compounds.~\cite{Fritsch05} Finally, Kondo screening is another alternative to quenching the magnetic entropy at higher temperatures without long range order.

Of all the CeTM$_4$In compounds, CePt$_4$In is most similar in structure and chemistry to CeIr$_4$In.  CePt$_4$In is a local moment system displaying heavy fermion behavior~\cite{Malik89,Pikul06}, with a Kondo temperature of $\sim$10~K and short range ordering occurring at 250~mK~\cite{Pikul08} which would favor some combination of the latter two scenarios for explaining the reduced entropy in CeIr$_4$In. If Ir hybridizes more strongly than Pt, this would favor the scenario where the magnetic anomalies are a result of impurity phases, and CeIr$_4$In is approaching mixed valent behavior akin to CeNi$_4$In. Consequently, it is currently inconclusive as to whether geometric frustration is playing a significant role in CeIr$_4$In. More measurements are needed to determine the origin of the physical properties in CeIr$_4$In.

\section{Discussion}

In retrospect, it may not be surprising to anticipate the discovery of CeIr$_{4}$In when the isostructural analog CePt$_{4}$In is known. However, all of the unknown Ce-Ir-In compounds computed in Table \ref{tab:results} contain closely related known isostructural analogs. Furthermore, it is remarkable that these relatively simple DFT calculations which neglect zero point motion, finite temperature, and electronic correlation effects, accurately predict the existence and lack thereof, for 10 out of the 13 ternary compounds and 15 of the 19 binary compounds investigated here in the Ce-Ir-In ternary phase diagram. In addition, for Ce$_{2}$IrIn$_{8}$ and CeIrIn$_{2}$ which are predicted to be unstable but are known, and Ce$_{2}$Ir$_{2}$In which is predicted to be stable but has yet to be found, the formation energies relative to competing phases are sufficiently small that we might expect that some of the neglected effects mentioned above might change the conclusion with regards to compound formation. We must then also acknowledge that given the small positive formation energy with respect to competing phases for CeIr$_{2}$In$_{4}$ and CeIr$_{2}$In$_{2}$, which is less than 56~meV/atom, it would not be surprising if either or both of these compounds are eventually synthesized.

In addition to the possible shortcomings of the DFT calculations mentioned earlier, we also recognize that the phase space of potential compounds which we are exploring is remarkably small. Not only have we selected a relatively small number of compositions, but we have also assumed a particular structure type for each of these compositions. In this respect, the addition of simulated annealing, genetic algorithms and other approaches~\cite{Woodley08} to this approach may aide in discovering additional structures not considered based on analogy.

The fact that arc melting Ce$_{2}$Ir$_{2}$In and annealing did not produce the anticipated structure does not necessarily mean that the compound is unstable.  However, it shows that Ce$_{2}$Ir$_{2}$In can not be a congruently melting compound and that the phase is either not stable at 800$^\circ$C or that a direct transformation from the competing phases may exhibit slow kinetics.  Alternative synthesis routes should be considered to stabilize the compound.  These could be as simple as using a different temperature annealing profile or using another synthesis route such as flux growth.~\cite{FiskFlux89} Thus, we anticipate that the effort in trying to synthesize a theoretically predicted compound by this method will be commensurate with the interest that the expected properties of that compound would generate.

Despite these potential complications, the fact that our combined computational and experimental approach demonstrates an accuracy of better than 80\% in identifying which compounds exist or not, proves that this can be a useful method for computationally identifying new stable structures.

\section{Summary}

We present a simple method combining density-functional calculations with arc-melting synthesis to efficiently investigate the structural stability of hypothetical compounds with the aim to improve the search for new compounds with desired structures and properties. As a proof of principle we applied our approach to the Ce-Ir-In phase diagram and predicted two previously unknown compounds, Ce$_{2}$Ir$_{2}$In and CeIr$_{4}$In, to be stable. Subsequent attempts to synthesize these compounds led to the discovery of CeIr$_{4}$In.  We expect that this approach will lead to the discovery of many more interesting compounds with specifically desired properties.

\section{Acknowledgments}

We thank Ben Ueland for assistance with the X-ray refinement, and Jon Lawrence for useful discussions.  Work at Los Alamos National Laboratory was performed under the auspices of U. S. Department of Energy, Office of Basic Energy Sciences, Division of Materials Sciences and Engineering and Theoretical Division under Contract No.\ DE-AC52-06NA25396 and supported in part by the Los Alamos Laboratory Directed Research and Development program.  We thank the Summer Student Program of the Center for Nonlinear Studies at Los Alamos National Laboratory for support.  The work at Cornell was supported by the National Science Foundation under Contract No.\ EAR-0703226 and through the Cornell Center for Materials Research NSF-IGERT: A Graduate Traineeship in Materials for a Sustainable Future under Contract No.\ DGE-0903653.  This research used computational resources of the National Center for Supercomputing Applications under Contract No.\ DMR050036 and the Computation Center for Nanotechnology Innovation at Rensselaer Polytechnic Institute.

\end{document}